\documentstyle[12pt]{article}

\begin{document}

\topmargin 0pt
\oddsidemargin 5mm

\setcounter{page}{1}
\vspace{2cm}
\begin{center}

{\bf 
FINITE SIZE EFFECTS FOR THE DILUTE COUPLING DERRIDA MODEL
}\\
\vspace{5mm}
{\
large Allakhverdyan A.E., Saakian D.B.
}\\
\vspace{5mm}
{\em Yerevan Physics Institute}\\
{Alikhanian Brothers St.2, Yerevan 375036, Armenia\\
Saakian @ vx1.YERPHI.AM}
\end{center}

\vspace{5mm}
\centerline{{\bf{Abstract}}}
We consider paramagnetic, spin-glass and ferromagnetic phases. At
$T=0$ model gives for the some values of connectivity (near the
critical) extremal suppression of finite size effects (decoding
error probability).

\vspace{3mm}
{\bf 1. Introduction}
\indent
Derridas's model \cite{BD} is the simplest among the spin-glasses.
It has solved in the first level of replica symmetry breaking
\cite{DG}.

May be this is the origin of unique feature-condition for the
appearance of ferromagnetic phase coincides with the constraint
of Shannon for the optimal coding.

This hypothesis \cite{NS} have proved for the general case in
\cite{DB},\cite{DB1}

The physical meaning of codes in the approach of Sourlas is this:

The $N$ numbers $\xi_i$ are given, taking values $\pm 1$.

By means of them we construct $Z$ numbers $\tau_i (Z\geq N)$,
then formulate some function of $N$ discrete
variables $\sigma_i$
\begin{equation}
\label{AA}
H(\sigma_i,\tau_j),
\end{equation}
which has as constants $Z$ couplings $\tau_j$,

There is a restriction, that single global minimum is
configuration
\begin{equation}
\label{AB}
{\sigma_i}={\xi_i}
\end{equation}

It is a simple to construct such function, but then in our
problem appears noise. Our couplings $\tau_i$ with probability
$\frac{1+m}{2}$ conserve their sign, and with probability
$\frac{1-m}{2}$ change it. Again there is a restriction,
that configuration (\ref{AB}) still is a single vacuum for the noisy
couplings $\tau_i$.

It is possible to do this procedure only for the some (good!) choice
of function $H$. For this case information theory gives
restriction \cite{IC}
\begin{equation}
\label{AC}
Z [\ln2-h(m)]\geq N\ln 2
\end{equation}
where for the entropy we have
\begin{equation}
\label{AD}
h(m)=-\frac{1+m}{2}\ln \frac{1+m}{2}-\frac{1-m}{2}\ln\frac{1-m}{2}
\end{equation}
Really our condition (single vacuum) is correct not with
probability 1, but a little less:
\begin{equation}
\label{AE}
1-a\,{\rm exp}[-E(m,N/Z)Z]
\end{equation}
The second member is just the decoding error probability.

On the language of statistical physics we will investigate
magnification at $T=0$
\begin{equation}
\label{AF}
<\xi_i\sigma_j>
\end{equation}
This expression gives probability, that our condition is fulo fild.
So decoding error probability is equivalent to finite size
effect in the expression of magnetization

In such manner we have coded our original information-$N$ numbers
$\xi_i$, to noisy number $\tau_j$. We can rederive original
values of $N$ numbers $\xi_i$, searching (by Monte-Karlo) minimum of
function $H(\sigma_i,\tau_j)$,

We want extremely suppress decoding error probability.

The information theory gives restriction for the maximal value of
function $E(m,R)$,

where
\begin{equation}
\label{AG}
R=N/Z
\end{equation}
is a rate of information transmission.

On the \cite{DB2} finite size effects have calculation at $T=0$ for
the full connectively case in Derrida model, which corresponds to
limit
\begin{equation}
\label{AH}
Z/N \rightarrow \infty \quad\quad\quad\quad m \rightarrow 1/2
\end{equation}
On the \cite{DB3} has investigated fine structure of ferromagnetic
phase, and has found some subphases, which differs each-others by
finite size effects.

In this work we are going to calculate finite size effects for
paramagnetic, spin-glass and ferromagnetic phase. For that
purposes we are using methods of \cite{CD}.

\section {Derivation of REM with weak connectivity}
\indent
Let us consider hamiltonian
\begin{equation}
\label{AJ}
H=-\sum_{(i_1 \cdots i_p)}
\,\,\,\tau_{i_1\cdots i_p}\,\sigma_{i_{1}}
\cdots\sigma_{i_p}
\end{equation}
where spins $\tau_i$, are quenched couplings $\tau_i$, taking values
$\pm 1$,

In our hamiltonian there are $C^P_N$ different choice of
$(i_1\cdots i_p)$ and consequently  $C^P_N$ different couplings
$\tau_{i_1\cdots i_p}$. We consider restriction
\begin{equation}
\label{AK}
\sum_{(i_1 \cdots i_p)} \quad(\tau_{i_{i}\cdots i_{p}})^2=\alpha N
\end{equation}
In the bounds of this restriction, which means an weak
correlation, all $\tau_j$ are distributed independently, with the
same probabilities
\begin{equation}
\label{AL}
\tau_{i_1 \cdots i_p}=0\quad {\rm with\, probability}\quad 1-\alpha
N/C_N^P
\end{equation}
\begin{equation}
\label{AQ}
\tau_{i_1 \cdots i_p}=\pm1\quad {\rm with\, probability}\quad
\frac{1\pm m}{2}\,\frac{\alpha N}{C^P_N}
\end{equation}
On \cite{CD} authors considered only conditions
(\ref{AL}),(\ref{AQ}). Our
restriction (\ref{AK}) simplifies calculations. As a results
finite size effects are changed only in paramagnetic phase

Let us  consider, as in \cite{CD}, common distribution of
$M$ energy levels.

We have $M$ configurations with the values of spins ${\sigma^\alpha
_i},\leq\alpha\leq M$ and values of energy $E_\alpha$.

Let as define
\begin{equation}
\label{AW}
P(E_1\cdots E_M)=<\delta(E_1-H(\sigma_i^1))\cdots\delta(E_M-
H(\sigma_i^M))>
\end{equation}

If it would be possible to calculate. Case with
\begin{equation}
\label{AZ}
M=2^N
\end{equation}
Then we could calculate
\begin{equation}
\label{AR}
<\ln Z(\tau,T)>\int\prod_{\alpha=1}^{2^N}\,dE_\alpha
P(E_1\cdots E_{2^N})\ln\sum^{2^N}_{\alpha=1}\ exp(- BE_\alpha)
\end{equation}
We have derived (\ref{AR}),using the fact that our system has
only $2^N$ levels of energy.

That's why we have considered $Z$ as a function of energy levels
$E_\alpha$ (instead of-as a function of couplings $\tau$). Then, due
to our choice (\ref{AK})-(\ref{AQ}),$P(E_1\cdots E_M)$ factorizates.

Using Fourier representation for $\delta$ function
\begin{equation}
\label{AT}
P(E_1\cdots E_M)=\int^{i\infty}_{-i\infty}\prod_{\alpha=1}^{2^N}
\,\frac{dE_\alpha}{2\pi}{\exp}\left[\sum^{M}_{\alpha=1}E_\alpha
\hat{E}_\alpha-\sum_{\alpha}\hat{E}_\alpha\sum_{{i_1}\cdots{i_p}}
\tau_{{i_1}\cdots{i_p}}
\sigma_{i_1}^\alpha \cdots \sigma_{i_p}^\alpha \right]
\end{equation}
Let us use standard representation
\begin{equation}
\label{AY}
\exp\,(a \sigma)=\cosh(a)\left[1+\tanh (a)\sigma\right],
\end{equation}
where $\sigma=\pm1$.

We can derive, expanding
\begin{equation}
\label{AU}
\prod_{{i_1}\cdots{i_p}}\left[ \cosh(\hat{E}_1)+
 m \sinh(\hat{E}_1) \right] \prod_{\alpha=2}^{M}\cosh(\hat{E}_\alpha)
\left[1+\tanh(\hat{E}_\alpha)\sigma_{i_1}^\alpha
\cdots \sigma_{i_p}^\alpha \right]
\end{equation}
the expression for the $P(E_1\cdots E_M)$
\begin{equation}
\label{AI}
P(E_1\cdots
E_M)=\int^{i\infty}_{-i\infty}\left\{\left[\cosh(\hat{E}_1)+
m\sinh(\hat{E}_1)\right]
\cosh(E_1)\cdots\cosh(E_M)\right\}^{\alpha N}
\prod_{\alpha=1}^M \frac{d E_\alpha}{2\pi}e^{\sum_{\alpha} E_\alpha
\hat{E}_\alpha}
\end{equation}
in the expression we neglect by effects of order
\begin{equation}
\label{AO}
\sum_{1\leq i_{1}\cdots<i_{p}\leq N}(\sigma_{i_1}^\alpha\,\sigma_{i_p}^
 B)\cdots (\sigma_{i_1}^\alpha\,\sigma_{i_p}^ B)/C^P_N
\end{equation}
We can consider this expression as a scalar product between 2
configurations).

Let us calculate (\ref{AO}) for the case, when only 1 turned spin.
it is ease to derive
\begin{equation}
\label{AP}
\left(C^P_{N-1}-C^{P-1}_{N-1}\right)/C^P_N=\frac{N-2P}{N}
\end{equation}
We see, that corrections disappear for the choice
\begin{equation}
\label{AS}
P=\frac{N}{2}
\end{equation}
When the number of turned spins is $\delta k $,
we have expression like
\begin{equation}
\label{AAA}
          \frac{1}{N}e^{-C\delta N}
\end{equation}
In future calculations we using expression (\ref{AI}) for
 $P(E_1\dots E_M)$. Then accuracy of one expression is polynomial
 or exponential (by $N$). We call this corrections
  as "finite $P$ corrections". Their investigation is a very
  hard work.

In the next section we calculating finite  size effects for the
model (\ref{AR}),(\ref{AI})using the formula from \cite{BD}
\begin{equation}
\label{AAB}
<\ln Z>=\Gamma^\prime(1)-\int\limits_{-\infty}^{\infty}\ln t e^{-\phi}d\phi
\end{equation}
where
\begin{equation}
\label{AAC}
e^{-\phi}=<e^{-tZ}>
\end{equation}

It is easy to derive, taking $Z=e^A$, equation
\begin{equation}
\label{AAD}
A=\Gamma^\prime(1)-\int\limits_{\infty}^{-\infty}u d\left[
e^{-e^{U+A}}\right]
\end{equation}
\section{Pharamagnetic Phase}
\indent
In this phase, as well as in the spin-glass case, we taking
$m=0$.

It is easy to derive
\begin{equation}
\label{AAF}
e^{-\phi}=\left[\int\limits_{-i\infty}^{i\infty}
\frac{dE_1dE_2}{2\pi}\, e^{-E_1E_2+\alpha N\ln \cosh E_1-t
e^{- B E_2}}\right]^{2^N}
\end{equation}

Let us make transformation
\begin{equation}
\label{AAG}
x=e^{- B E_2},\quad E_1= B E
\end{equation}
 After this easy to derive
 \begin{eqnarray}
\label{AAH}
e^{-\phi}&=&\left[\int\limits_{-i\infty}^{-i\infty}
\frac{dE}{2\pi} \int\limits_{0}^{\infty}
e^{-x+(E-1)\ln x -UE+\alpha N\ln\cosh (B E)}\right]^{2^N}\nonumber\\
&=&\left[\int\limits_{-i\infty}^{-i\infty}
\frac{dE}{2\pi}\Gamma (E)
e^{-UE+\alpha N\ln\cosh (B E)}\right]^{2^N}
\end{eqnarray}

In this integral integration loop passed pole zero from right
side.

For the integration we must lift integration loop righter,
until saddle point, defined by equation
\begin{equation}
\label{AAJ}
\tanh(BE)=\frac{U}{\alpha N B}
\end{equation}

Doing this we must take care for contribution of intersected
poles.

To calculate thermodynamic limit it is enough to take account
poles  $0,-1$
\begin{equation}
\label{AAK}
e^{-\phi}\simeq\left[1-e^{U+\alpha N\ln\cosh B}
\right]^{2^N}\simeq {\rm exp}\left[-e^{U+N\ln 2+\alpha N\ln\cosh B}
\right]
\end{equation}
Using formula (\ref{AAB}), we derive
\begin{equation}
\label{AAZ}
<\ln Z>\approx N\ln 2+\alpha N\ln\cosh B
\end{equation}

Let us take in to account finite size effect. The saddle point
gives
\begin{eqnarray}
\label{AAX}
e^{-\phi}&\simeq&\left[1-e^{U+\alpha N\ln\cosh B}
+\Gamma(E)e^{-EU+\alpha N\ln\cosh (B E)}\right]^{2^N}\nonumber\\
&\simeq& \left[e-e^{U+\alpha N\ln\cosh B}+\Gamma(E)e^{-(E+1)U
+\alpha N\ln\cosh (B E)\/cosh  B}\right]^{2^N}\nonumber\\
&=& e^{-U+\alpha N\ln\cosh B+ N\ln 2+\varphi(U)}
\end{eqnarray}

So we found connection between $\varphi$ and $U\equiv\ln t$
\begin{equation}
\label{AAV}
\ln\varphi=U+U_0+\varphi(U)
\end{equation}
where
\begin{eqnarray}
\label{AAN}
U_0&=&-(N\ln 2+\alpha N\ln\cosh B)\nonumber\\
\varphi(U)&=&\Gamma(E){\rm exp}\left[-(E+1)U+\ln\cosh (B E)/\cosh B\right]
\end{eqnarray}
Let us express $U$ by $\ln\phi$
\begin{equation}
\label{ BA}
u\approx-\ln\phi-U_0-\varphi(U_0)
\end{equation}
With such accuracy
\begin{eqnarray}
\label{ BS}
<\ln Z>&=&\int\limits^{\infty}_{0}\ln t (e^{\phi} d\phi)\nonumber\\
&=&\alpha N\ln\cosh B+N\ln
2+\Gamma(E)e^{N(E+1)\ln\delta+\alpha\ln\left[\cosh (BE)(\cosh
B)^E\right]}
\end{eqnarray}
In the last expression we have neglected preexponent besides
$\Gamma(E)$. The next subphase appears, when saddle point $E$
coincides with point-2:
\begin{equation}
\label{ BD}
\tanh(2 B)=\frac{\alpha\ln\cosh B+\ln 2}{\alpha B}
\end{equation}
The equation about subphase of paramagnetic phase demands more
careful discussion.
\section{\bf Spin-glass phase}
\indent

In this case it is enough to intersect only one pole $E=0$

After transformation we have
\begin{equation}
\label{ BF}
e^{-\phi}=\left[1+\frac{1}{2\pi}\int\limits_{-i\infty}^{i\infty+E_0}
d E \Gamma(E)e^{-EU+\alpha N\ln\cosh (B E)}\right]^{2^N}
\end{equation}
\begin{equation}
\label{ BG}
e^{-\phi}\equiv{\exp}\left[\frac{-|\Gamma(E_0)|}{h\sqrt{2\pi}}
{\exp}(-E_0U+\alpha N\ln\cosh (B E_0)+N\ln 2)\right]
\end{equation}
where $h=\sqrt{\alpha N B}$ and for $E_0$ we have equation
\begin{equation}
\label{ BH}
\tanh (B E_0)=U/\alpha N B
\end{equation}

Let us express again $U$ by $\ln\phi$
\begin{equation}
\label{ BJ}
\ln\phi \approx E_0U+\alpha N\ln\cosh (B E_0)+N\ln
2-\frac{1}{2}\ln N
\end{equation}
\begin{equation}
\label{ BK}
U\approx\frac{1}{E_0}\ln\phi+\frac{\alpha
N}{E_0}\ln\cosh (B E_0)+\frac{N\ln 2}{E_0}-\frac{1}{2}\frac{\ln
N}{E_0}
\end{equation}

In this formula we deal with expressions like $ B E_0$ and
$U/ B$.
\begin{equation}
\label{NA}
\tanh (B E_0)=\frac{\ln\cosh B E_0+\ln
2/\alpha}{ B E_0}-\frac{\ln N}{2\alpha N B
E_0}+\frac{\ln\phi}{\alpha N B E_0}
\end{equation}

If we neglecting the last terms in (\ref{NA})then we have
\begin{equation}
\label{NS}
E_0=\frac{- B_ c}{ B}
\end{equation}

Where at $ B_ c$ disappears entropy. For corrections we have
$(x=B E_0)$
\begin{eqnarray}
\label{ND}
(x\tanh x-\ln\cosh x-ln 2\alpha)=-\frac{\ln N}{2\alpha
N}\nonumber\\
x=x_0-\frac{(\cosh x_0)^2}{x_0}\frac{\ln N}{2\alpha N}\nonumber\\
U=\alpha N B\tanh x=\alpha N B\tanh x_0-\frac{1}{2}\ln N
\frac{ B}{ B_c}
\end{eqnarray}
For the free energy we receive
\begin{equation}
\label{NF}
<\ln Z>=\alpha N
 B\tanh B_c-\frac{1}{2}\frac{ B}{ B_c}\ln N
\end{equation}
So at phase transition point free energy has jump
\begin{equation}
\label{NG}
-\frac{1}{2}\ln N
\end{equation}
as for the case fully connected model
\section{\bf Ferromagnetic Phase}
\indent
For $e^{-\phi}$ we have expression
$$e^{-\phi}\approx G(U)\tilde{G}(U)^{2^N-1}$$
where
\begin{equation}
\label{NH}
\tilde{G}(U)=\left\{
 \begin{array}{ll}\displaystyle 1-\frac{\cosh B E_0
|\Gamma(E_0)|}{2\sqrt{\pi}}e^{-E_0U+\alpha
N\ln\cosh (B E_0)}&, E_0<0;\\
\displaystyle
\frac{\cosh B E_0 |\Gamma(E_0)|}{2\sqrt{\pi}}e^{-E_0U+\alpha
N\ln\cosh (B E_0)}&, E_0>0
\end{array}\right.
\end{equation}
and for $E_0$ we have equation
\begin{equation}
\label{NJ}
\tanh( B E_0)=\frac{U}{\alpha N B}
\end{equation}
Similarly for $G(U)$
\begin{eqnarray}
\label{NK}
&&G(U)=\\
&&=\left\{
\begin{array}{ll}
\displaystyle 1-\frac{(\cosh( B E_1)-m \sinh(  B E_1))}
{2\sqrt{\pi}\sqrt{1-m^2}}|\Gamma(E_1)| e^{-E_1 U+\alpha N\ln
[\cosh(B E_1)-m \sinh(B E_1)]}&,E_1<0\\
\displaystyle
\frac{\cosh B E_1-m \sinh  B E_1}{2\sqrt{\pi}\sqrt{1-m^2}}
|\Gamma(E_1)| e^{-E_1 U+\alpha N\ln[\cosh(B E_1)-m \sinh
 (B E_1)]} &, E_1>0.
\end{array}
\right. \nonumber
\end{eqnarray}
and for $E_1$
\begin{equation}
\label{NL}
\frac{m \tanh  B E_1-m}{1-m\tanh E_1}=x, \quad
x=\frac{U}{\alpha N  B}
\end{equation}
Let us define function
\begin{eqnarray}
\label{MA}
f(x)&=&\frac{1+x}{2}\ln\frac{1+x}{1+m}+\frac{1-x}{2}\ln\frac{1-x}{1+m}\\
g(x)&=&\frac{1+x}{2}\ln\frac{1+x}{2}+\frac{1-x}{2}\ln\frac{1-x}{2}
+\ln 2\\
\end{eqnarray}
Simple calculations give
\begin{eqnarray}
\label{MS}
-E_0 U+\alpha N\ln\cosh (B E_0)&=&-\alpha N g(x)\\
-E_1 U+\alpha N\ln\cosh  (B E_1)&=&-\alpha N f(x)
\end{eqnarray}

We wont to calculate corrections
\begin{equation}
\label{MD}
<\ln Z>-\alpha N B m=\int G(U)\left[1-\tilde{G}(U)^{2^N-1}\right]
\equiv I
\end{equation}
It is convenient to break integration interval
\begin{equation}
\label{ME}
I=\int\limits^{-U_0}_{-\infty}+\int\limits^{-U_1}_{-U_1}
+\int\limits^{\infty}_{-U_1}
\end{equation}
Where
\begin{equation}
\label{MF}
U_0=\alpha N B m
\end{equation}
and
\begin{equation}
\label{MG}
U_1=\alpha N B x_1
\end{equation}
The value of $x_1$ defined from equation
\begin{equation}
\label{MH}
\alpha g{x_1}=\ln 2
\end{equation}

In the first integral we see the condition of dominance of ferromagnetic
free energy under spin-glass expression:
\begin{equation}
\label{MJ}
\alpha g(m)>\ln 2
\end{equation}
The main contribution to the finite size correction comes from second
integral.

We have two subphases. In the first out to saddle point is in the
integral of integration, in the second one-out side.
\begin{equation}
\label{MK}
I\sim {\rm exp}\left[-\alpha N f(-x_1)\right], \quad x_1>x_2
\end{equation}
and
\begin{equation}
\label{ML}
I\sim {\rm exp}\left[-\alpha N f(-x_2)-\alpha N g (x_2)+
N \ln 2\right], \quad x_1>x_2
\end{equation}
where for $x_2$ we have
\begin{equation}
\label{VV}
x_2=\frac{1-\sqrt{1-m^2}}{m}
\end{equation}

Our expression (\ref{MK}),(\ref{ML}) coincide with known results
of information theory
\section{\bf Summary}
\indent
We have calculated finite size effects for Derrida model with
weak connectivity.

In the spin-glass phase we recived logarithmic corrections, in
paramagnetic-expotential.

In ferromagnetic phase at $T=0$ our results coincide with
information theory (the results of random coding).That codes give
for the value of connectivity near the critical extremal possible
suppression of decoding error probability. For strong
connectivity the result is unknown for optimal coding.

Unfortunally known boundaries from information theory show, that
Derridas model does not give extremal suppression for the case or
high values of connectivity (weak velocity). To clarify
situation, we need consider finite $P$ corrections.

In the work \cite{PR},\cite{HN} has suggested to consider
\begin{equation}
\label{VS}
<|\sigma_i(\tau, T_N)|>_\tau
\end{equation}
instead of magnetization at $T=0$, where
$T_N=\frac{1}{2}\ln\frac{1+m}{1-m}$

Neglecting finite $P$ corrections we received the same result for
the exponent, as for the case $T=0$

We would like to thank Yev.Harutunyan for consultations during
the work.

We are also grateful L. Basalyga, R.Dobrushin,  B.Derrida,
Sh.Rouhani. N.Sourlas for useful remarks.

This work was supported
by NATO linkage grant LG 9303057,
ISF grant MVMOOO and German ministry of
research and technology grant 211-5231. D.B. Saakian thanks 
Ivan Kostov for hospitality in Saclay.

\newpage

\setcounter{equation}{0}
\appendix{{\bf Appendix A}}
\renewcommand{\theequation}{A.\arabic{equation}}

\indent
In this appendix we show, how transform our results for the case
of full connectivity and derive results of \cite{BD}.

For this purposes we introduce coupling $J>0$, so we consider
\begin{eqnarray}
\label{CA}
P(j_{i_1\dots i_P})&=&\left(1-\frac{\alpha N}{2 C_N^P}\right)
\delta (j_{i_1\dots i_p})\nonumber\\
&+&\frac{\alpha N}{2C_N^p}\left[\frac{1+m}{2}
\delta(j_{i_1\dots i_p}-J)+\frac{1-m}{2}\delta(j_{i_1\dots i_p}+J)
\right]
\end{eqnarray}
 For this purposes it is just enough to replace
 $ B\rightarrow BJ$, $ B_ c \rightarrow  B_cJ $

It is easy to cheek \cite{IC}, that transition to full connectivity
means
\begin{equation}
\label{CB}
J\rightarrow J \alpha^{-1/2} ,\quad\quad\alpha\rightarrow\infty
\end{equation}

For us it means
\begin{equation}
\label{CD}
e^{-\phi}=\left[\int\limits_{-i\infty}^{i\infty}
\frac{dE}{2\pi}\Gamma(E)e^{-Eu+\frac{N(JBE)^2}{4}}
\right]^{2^N}
\end{equation}

We consider the case $m=0$.

Different expressions for $e^{-\phi}$ depends on the poles,
intersected by our loop for saddle point intergation at
$E=2 U/\lambda^2$ and the last
depends on the temperature. It is easy to derive (\ref{CD})
directly.

\end{document}